\documentclass[12pt]{article}
\usepackage[dvips]{graphicx,color}
\usepackage[dvips]{psfrag}

\usepackage{cite}

\usepackage{t1enc,a4wide}
\usepackage{amsmath,amssymb,euscript,wrapfig}

\makeatletter
\def\fmslash{\@ifnextchar[{\fmsl@sh}{\fmsl@sh[0mu]}}
\def\fmsl@sh[#1]#2{%
  \mathchoice
    {\@fmsl@sh\displaystyle{#1}{#2}}%
    {\@fmsl@sh\textstyle{#1}{#2}}%
    {\@fmsl@sh\scriptstyle{#1}{#2}}%
    {\@fmsl@sh\scriptscriptstyle{#1}{#2}}}
\def\@fmsl@sh#1#2#3{\m@th\ooalign{$\hfil#1\mkern#2/\hfil$\crcr$#1#3$}}
\makeatother

\long\def\symbolfootnote[#1]#2{\begingroup%
\def\thefootnote{\fnsymbol{footnote}}\footnote[#1]{#2}\endgroup}

\begin{document}
\begin{titlepage}
\begin{flushright}
SI-HEP-2007-21 \\[0.2em]
{16.04.2008, revised 10.07.2008}
\end{flushright}

\vspace{1.2cm}
\begin{center}
\Large\bf\boldmath
Is there a non-Standard-Model contribution \\ in non-leptonic $b\to s$ decays?  
\unboldmath 
 \end{center}

\vspace{0.5cm}
\begin{center}
{\sc Thorsten Feldmann,
 Martin Jung, Thomas Mannel} \\[1.5em]
{\sf Universit\"at Siegen, Fachbereich Physik\\ 
D-57068 Siegen, Germany.} \\[1em]
\end{center}

\vspace{0.8cm}
\begin{abstract}
\vspace{0.2cm}\noindent
Precision measurements of branching fractions and CP asymmetries
in non-leptonic $b \to s$ decays reveal certain ``puzzles'' when
compared with Standard Model expectations based on
a global fit of the CKM triangle and
general theoretical expectations. Without reference
to a particular model, we investigate to what extent the 
(small) discrepancies observed in $B \to J/\psi K$, $B \to \phi K$
and $B \to K\pi$ may constrain new physics in $b \to s q\bar q$
operators. In particular, we compare on a quantitative level the
relative impact of different quark flavours $q=c,s,u,d$.
\end{abstract}

\end{titlepage}

\section{Introduction}
Exclusive non-leptonic $B$ meson decays remain a challenge to 
theory. While semi-leptonic $B$ decays are well described within the heavy-mass
expansion and allow for a rather precise determination of the CKM matrix elements 
$|V_{cb}|$ and $|V_{ub}|$, exclusive non-leptonic decays still cannot be described 
at a similar level of precision.
The methods that have been proposed so far are based on the flavour symmetries
of QCD 
\cite{Gronau:1990ka,Nir:1991cu,Fleischer:1995cg,Neubert:1998re,Grossman:2003qp,Buras:2004ub,Datta:2004re}, 
the factorization of QCD dynamics in hadronic matrix elements
\cite{Beneke:2000ry,Beneke:2001ev,Keum:2000wi,Bauer:2005kd,Williamson:2006hb}, 
or combinations thereof \cite{SR,Lipkin:1998ie,Matias:2001ch}. The level of precision that one expects
from these methods is typically of the order of tens of percent, 
and thus -- except for a few ``gold-plated'' observables --
it will in general be hard to pin down an effect from new physics (NP) in these decays. 
Still, from the experimental side, the B-factories have collected sufficient information 
on decay widths and CP asymmetries to allow for global fits of the 
Standard Model (SM) parameters, in particular of the elements of the
Cabibbo-Kobayashi-Maskawa (CKM) matrix \cite{Charles:2004jd,Bona:2006ah}.

The agreement between the standard theory and experimental data is 
overall satisfactory, however, in some cases small tensions appear. 
In the present paper we focus on the $|\Delta B| =|\Delta S| = 1$ 
decay modes $B \to J/\Psi K$, $B \to \phi K$ and $B \to K \pi$, which enter 
some of the present-day ``puzzles''. Taking the experimental results at
face value, we pursue the hypothesis that these ``discrepancies''
with the SM calculations are due to non-standard effects \cite{Peskin:2008zz}.  
We adopt a model-independent parameterization in terms of isospin
amplitudes, where we allow for additional contributions
from generic NP operators. The moduli, as well
as the strong phases of the additional terms are then fitted
to experimental data on decay widths and CP asymmetries. The new
{\em weak phase} will generally remain undetermined due to
reparameterization invariance, as long as
we do not attempt to fix the hadronic SM matrix elements. 
In the case of $B \to K\pi$ decays we make use of
additional theoretical input from
the QCD-improved factorization approach (QCDF) \cite{Beneke:2001ev}.

Our paper is organized as follows:
In the next section we point out the tensions of the SM fit 
with present data, and give arguments for the way we are going to re-fit 
the experimental data including generic NP contributions. 
In the following section we discuss
the results of the fits for the individual decay modes and present our conclusions in Section~\ref{sec:concl}.

\section{Phenomenology}

\begin{figure}[t!hb]
\begin{center}
\parbox[c]{0.5\textwidth}{
\includegraphics[width=0.45\textwidth]{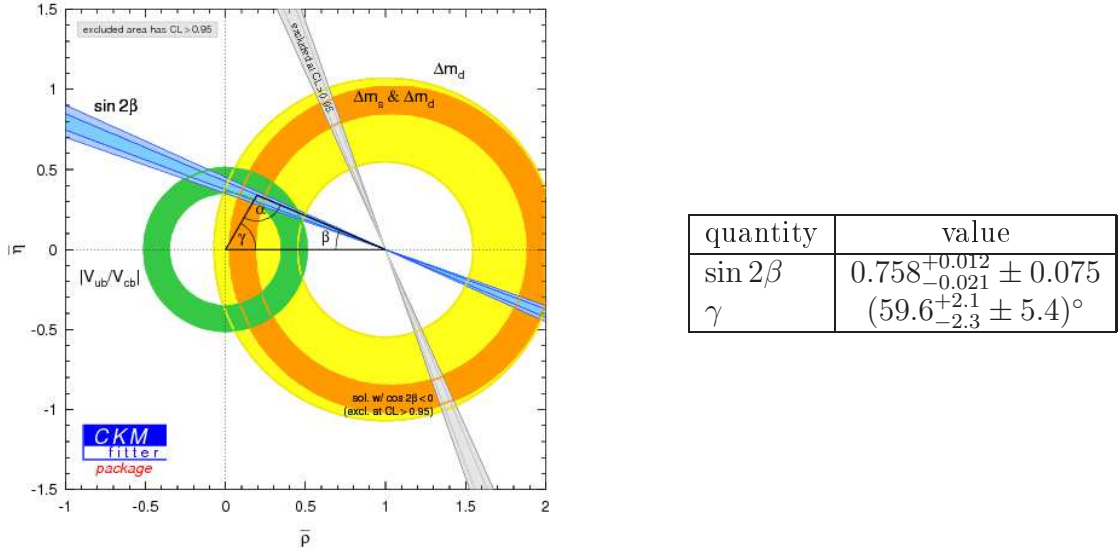}
} \qquad 
{\begin{tabular}{|l|c|}\hline
 quantity & value\\\hline
 $\sin2\beta$ & $0.758 ^{+0.012}_{-0.021} \pm 0.075$\\
 $\gamma$ & $(59.6^{+2.1}_{-2.3}\pm 5.4)^\circ$ \\
\hline
\end{tabular}}
\vspace{0.8em}
\parbox{0.9\textwidth}{
\caption{\label{figNNPs2bga}\small
Global fit to CKM parameters from $\Delta m_d$, $\Delta m_s$ and
$|V_{ub}/V_{cb}|_{\rm excl.+incl.}$. Left: Confidence levels in the $\bar\eta-\bar\rho$\/ plane.
Right: Fitted values for CKM parameters, where the first
error is treated as Gaussian, and the second error is treated as flat.
}}
\end{center}
\end{figure}


\subsection{Tensions with the Standard Model?} 
In the following we give a brief discussion of the present situation
for the $B$ physics observables that we are going to consider,
where the standard model displays some 
tension with the data:
\begin{itemize} 

\item
The first point concerns the global fit of the CKM unitarity triangle. Here a 
small mismatch appears between the value of the CKM angle $\beta$ obtained from the 
direct measurement of the time-dependent CP asymmetry in $B \to J/\psi K_S$ 
and the indirect determination of the same angle from the mass differences in
the neutral $B$\/-meson systems, $\Delta m_d / \Delta m_s$, 
in combination with the measurement of $|V_{ub}/V_{cb}|$ from semi-leptonic
decays \cite{Bona:2006ah,Lacker:2007me,CKMprivate,Yao:2006px,Abulencia:2006ze,Kowalprivate}. 
In fact, using the values from \cite{Lacker:2007me}, 
we find for the latter case  $\sin 2 \beta = 0.758 ^{+0.012}_{-0.021} \pm 0.075_{\rm flat}$
(see Fig.~\ref{figNNPs2bga}), 
while the direct determination using $B \to J/\Psi K_S$ yields 
$\sin 2 \beta = 0.678 \pm 0.025$ \cite{Barberio:2006bi}.
However, the significance
of this effect depends strongly on the estimates of the theoretical uncertainties,
e.g.\ in the determination of $|V_{ub}|$, and can certainly not be taken as a clear
evidence for a non-standard effect.

\item
The second puzzle arises from the time-dependent CP asymmetry in modes like $B \to \phi K_S$ 
which in the SM again yields a determination of $\sin2\beta$, although 
with less precision. The value for $\beta$ obtained from fits to several
$b \to s \bar{s} s$ penguin modes\footnote{%
Following the arguments given by HFAG \cite{Barberio:2006bi}, we do not consider
the $\sin2\beta$ value extracted from 
$B^0 \to f_0 K_S$ for our discussion, due to the
highly non-Gaussian error implied by the BaBar measurement \cite{Aubert:2007vi}.
}
does not agree with the value from
the CP asymmetry in $B \to J/\Psi K_S$ \cite{Barberio:2006bi}. 
While part of the discrepancy may be due to
not well understood hadronic effects, it is at least curious that the bulk of 
decay modes involving the $b \to s\bar{s}s$ penguin systematically yields a lower value for 
$\sin 2 \beta$ than the one obtained from $B \to J/\Psi K_S$
(see also \cite{Sinha:2006rj}).

\item
Finally, the theoretical predictions for $B \to K\pi$ decay widths and CP asymmetries 
are not always in very good agreement with the data. Within the QCD factorization approach
the discrepancy with the data can be brought to an ``acceptable'' level 
(except for, perhaps, the differences of CP asymmetries $\Delta A$,
see the discussion in Sec.~\ref{sec:npkpi} below)
by assuming particular
scenarios within the hadronic parameter space, including undetermined $1/m_b$ corrections \cite{Beneke:2003zv}. On the other hand, analyses based on $SU(3)$ flavour symmetry for hadronic
matrix elements typically have found tensions of the order of $(2-3) \, \sigma$
\cite{Buras:2004ub,Bauer:2005kd,SR,Charles:2004jd,Mishima:2004um,Kim:2005jp,Baek:2006ti,Baek:2007yy},
depending on additional assumptions about hadronic matrix elements. 
It should be noted that the tensions
related to the branching fractions have decreased since the inclusion of
electromagnetic corrections in the experimental analysis (for a recent update of the 
discussion see, for instance, \cite{Fleischer:2007mq}).

\end{itemize}

Let us, for the moment, take these tensions between theoretical expectations and
experimental data at face value: Assuming that they are not due to 
enormous deviations from the factorization approximation to hadronic matrix elements, 
we may try to localize in which part of the effective weak Hamiltonian we have to look 
for NP effects. 

A first possibility are non-standard contributions in the 
charged $b \to u$ current which determines $|V_{ub}|$. 
However, it is generally believed to be unlikely 
that these tree-level processes contain sizeable NP effects. Likewise, 
the theoretical description of QCD dynamics in semi-leptonic decays is 
fairly well under control, and hence we will not consider this possibility here. 

A second explanation could be a non-standard contribution in the mixing
phase of the $|\Delta B| = 2$ part of the effective Hamiltonian, which shifts the 
observed  $\sin 2 \beta$ to smaller values. Such a scenario corresponds to a 
generalization  of Wolfenstein's ``superweak interaction''
\cite{Wolfenstein:1964ks}. Obviously, it cannot 
explain the differences in the $\sin2\beta$ measurements from $b\to c\bar c s$
and $b \to s\bar ss$ modes.

The third scenario, which is the one we are going to expand on in this work,
is the possibility to have an additional contribution in the
$|\Delta B| = |\Delta S|=1$ part of the effective Hamiltonian.
Evidently, the inclusion of such terms can explain the findings 
in $B \to J/\Psi K_S$ and $B \to \phi  K_S$, as well as in $B \to K\pi$.
When fitted to experimental data, the values for the 
NP contributions, relative to the leading
hadronic amplitudes in the SM, can be as large as about 30\%.
If NP is the explanation for the tensions in non-leptonic $b \to s$ transitions,
structures beyond minimal-flavour violation
(MFV \cite{D'Ambrosio:2002ex,Ciuchini:1998xy,Buras:2000dm}) 
are favoured, mainly because the deviations in the $B \to \phi K$
CP asymmetries point towards an independent NP phase, but also because
the constraints on contributions from different flavours $q$ in $b \to s q\bar q$
generally can be rather different in size.


\section{Fit of new physics contributions to experimental data}

\subsection{$|\Delta B| = |\Delta S|=1$ transitions}

Using the unitarity of the CKM matrix, the SM operator basis 
for non-leptonic $b \to  s$ transitions can be written as \cite{Buchalla:1995vs}
\begin{equation}
 H_{\rm SM}^{|\Delta B|=|\Delta S|=1} = 
\frac{G_F}{\sqrt2} \,
V_{cb} V_{cs}^* \left( C_{1,2} \, O_{1,2}^{(c)} + \sum_{i\geq3} C_i \,O_i \right)
+\frac{G_F}{\sqrt2} \,
V_{ub} V_{us}^* \left( C_{1,2} \, O_{1,2}^{(u)} + \sum_{i\geq3} C_i \, O_i \right) \,,
\label{HSM}
\end{equation}
where $O_{1,2}^{(q)}$ are the current-current operators, $O_{3-6}$ the strong penguins,
$O_{7-10}$ the electroweak penguins, and $C_7^\gamma$, $C_8^g$ the electromagnetic
and chromomagnetic operators, respectively.
At low energies, the effect of NP in $|\Delta B| = |\Delta S|=1$ transitions will be parameterized by new dimension-six operators.
In the following we shall focus on generic four-quark operators
of the type $b\to s q\bar q$ with $q=(b),c,s,u,d$,
where the Dirac and colour structure will not be specified.
In order to quantify the possible size of
NP contributions, we will always assume the dominance of one particular
flavour structure, and parameterize the corresponding correction to
the SM decay amplitudes in a model-independent way.

\subsection{Statistical framework}

The parameter space for the NP amplitudes is explored using the
CKMfitter package \cite{Hocker:2001xe}.
Here the amplitude parameters
are treated as fundamental theoretical quantities, and the statistical
analysis provides the relative likelihood for a given point in parameter space
(corresponding to model-dependent ``metrology'' in the CKMfitter jargon). 
Other theoretical parameters, like hadronic uncertainties from SM physics, are 
treated using the Rfit-scheme, where the corresponding $\chi^2$\/-contribution is 
set to zero within a ``theoretically acceptable'' range, and set to infinity outside.
We will sometimes apply the same approach to implement
additional theoretical constraints/assumptions 
on the amplitude parameters, in
order to suppress ''unphysical'' solutions.

\subsection{Analysis of $B \to J/\Psi K$} 

For $B \to J/\Psi K$ decays the contribution of the
second term in the weak effective Hamiltonian~(\ref{HSM}) is small because of two effects:
\begin{itemize}
 \item Cabibbo suppression: $|V_{ub} V_{us}^*|/|V_{cb} V_{cs}^*| \sim \lambda^2 \ll 1$ 
 \item Penguin suppression: (i) The operators $O_{1,2}^{(u)}$ do not contain charm quarks, 
       and the hadronic matrix elements $\langle J/\psi K| O_{1,2}^{(u)}|B\rangle$ 
       are suppressed. (ii) The coefficients of the loop-induced penguin operators $C_{i\geq 3}$ 
       are small with respect to the tree coefficients $C_{1,2}$.
\end{itemize}
Furthermore, the electroweak penguin operators can be neglected compared to
the strong penguin operators. Consequently, in the SM the $B \to J/\psi K$ 
decay amplitude is expected to be completely 
dominated by
\begin{equation}
{\cal A}_0 (\bar B \to J/\psi \bar K) = \frac{G_F}{\sqrt{2}} \, V_{cb} V_{cs}^* \,
 \langle J/\psi \bar K | C_{1,2} \, O_{1,2}^{(c)} + \sum_{i=3}^{6} C_i O_i^{(c)}|\bar B \rangle 
\end{equation} 
where $\bar B = \{ \bar B^0_d,\, B^- \}$, and we projected out 
the leading $[\bar s b \bar c c]$ component in every operator,
$O_i \to O_i^{(c)}$.
In particular, the amplitude is dominated by a single weak phase, and consequently
the time-dependent CP asymmetry in $B^0 \to J/\psi K_S$ is 
completely determined by the $B^0 - \bar B^0$
mixing amplitude, involving the CKM angle $\beta$. Corrections from the sub-leading
operators have been estimated by perturbative methods at the $b$-quark scale,\footnote{The 
authors of \cite{Boos:2004xp} only considered the effect of $O_{1,2}^u$. In \cite{Li:2006vq}
important contributions from $C_{3-6}$ have been included as well.}
and found to give effects of the order of $10^{-3}$, only \cite{Boos:2004xp,Li:2006vq}.
Long-distance penguin contributions have been estimated on the basis of experimental data
to be not larger than $10^{-2}$ \cite{Ciuchini:2005mg}.
\begin{table}[t!!!t]
\begin{center}
\parbox{0.9\textwidth}{
\caption{\small Partial widths \cite{Yao:2006px}
 and CP asymmetries \cite{Barberio:2006bi}
 for $B\to J/\Psi K$.
\label{tab:psi}}\vspace{0.8em}}
{\begin{tabular}{|l|c|}\hline
 $\eta_{\rm CP} \, S_{J/\psi K_S}$ & $ -0.678 \pm 0.026$ \\ \hline
 $C_{J/\psi K_S}$ & $ \phantom{+}0.012 \pm 0.020$ \\
 $A_{\rm CP}(J/\psi K^-)$ & $\phantom{+}0.015 \pm 0.017$ \\
\hline
$\Gamma(B^-\to J/\Psi K^-)$ & $(6.13\pm 0.22) \cdot 10^{-4}\, {\rm ps}^{-1}$\\
$\Gamma(\bar{B}^0\to J/\Psi \bar{K}^0)$ & $(5.71 \pm 0.22)\cdot 10^{-4} \, {\rm ps}^{-1}$ \\
\hline
\end{tabular}}
\end{center}
\end{table}
Furthermore, the dominating operators in the SM decay amplitude conserve strong
isospin ($\Delta I=0$), and therefore do not induce differences between the charged
and neutral $B$ decays into $J/\psi K$. The present experimental data is
summarized in Table~\ref{tab:psi}. We note that the central value for 
$S_{J/\psi K_S}$ differs from the indirect determination for $\sin2\beta$
in Fig.~\ref{figNNPs2bga}, but the two values are consistent within the errors.
The discrepancy becomes slightly more pronounced, if one takes into account
the inclusive measurement for $|V_{ub}|$ only, which gives
$$
  \sin2\beta = 0.821^{+0.024}_{-0.046} \pm 0.068_{\rm flat} \qquad \mbox{(using $|V_{ub}|_{\rm incl.}$ from 
 \cite{Lacker:2007me}).}
$$
Similarly, the central values for the
observed isospin-breaking in the CP asymmetries
($C_{J/\psi K_S}$ vs.\ $-A_{\rm CP}(J/\psi K^-)$) and partial widths
differ from zero.

Allowing for generic NP contributions with \emph{one}
weak phase $\theta_W$, the amplitudes can be written as
\begin{eqnarray} 
{\cal A} (B^-\to J/\psi K^-) 
 &=& {\cal A}_0 (\bar B \to J/\psi K) 
 \left[ 1 + r_0 \, e^{i \theta_W}  e^{i \phi_{0}} -
    r_1 \, e^{i \theta_W}  e^{i \phi_{1}} \right] \,,
\cr
{\cal A} (\bar B_d \to J/\psi \bar K^0) 
 &=& {\cal A}_0 (\bar B \to J/\psi K) 
 \left[ 1 + r_0 \, e^{i \theta_W}  e^{i \phi_{0}} + 
    r_1 \, e^{i \theta_W}  e^{i \phi_{1}} \right] \,,
\label{NewAmp}
\end{eqnarray}
where we have separated the contributions to transitions with $\Delta I=0$
(i.e.\ tree-level matrix elements with $b \to s c\bar c$ operators 
 or long-distance strong penguins with $b\to s (u\bar u + d\bar d)$ 
 or $b \to s s\bar s$ operators)  and
$\Delta I=1$ 
(annihilation topologies with $b \to s u\bar u$ or $b \to s d\bar d$). 
We introduced the absolute values $r_0$,  $r_1$ and strong phases
$\phi_{0}$,  $\phi_{1}$ for the hadronic matrix elements
associated with the corresponding NP operators, relative to
the leading SM amplitude.

\subsubsection{Fit with $\Delta I=0$ only (new physics in $b \to s c \bar c$)}

Among the $\Delta I=0$ and $\Delta I=1$ operators we expect the 
$b\to s c \bar c$ term to give the dominating contributions to 
$B \to J/\psi K$ decays, because it has (unsuppressed) tree-level matrix 
elements with the hadronic final state. Therefore, let us first
assume that $b\to s c \bar c$ gives the only relevant NP contribution
in (\ref{NewAmp}) which amounts to setting $r_1$ to zero, while $r_0$
should be of the order $m_W^2/\Lambda_{\rm NP}^2$.
Then, the isospin breaking between charged and neutral
$B$ decays is not affected, and should not be part of the fit. 
We are thus left with the time-dependent CP asymmetries, defined as 
in \cite{Charles:2004jd}
\begin{equation} \label{CPA}
A_{\rm CP}(f,t) := 
\frac{{\rm BR}[\bar B^0 \to f](t) - {\rm BR}[ B^0 \to \bar f](t)}
     {{\rm BR}[\bar B^0 \to f](t) + {\rm BR}[ B^0 \to \bar f](t)}
 := - C_f \cos(\Delta m \, t) + S_f  \sin(\Delta m \, t) 
\end{equation}
and the direct CP asymmetry $A_{\rm CP}^{\rm dir}(B^- \to J/\psi K^-)=-C_{J/\psi K_S}$.
Including the contribution from $r_0$ in (\ref{NewAmp}),
we obtain
\begin{eqnarray}
C_{J/\psi K_S} &=& - A_{\rm CP}^{\rm dir}(B^- \to J/\psi K^-)\nonumber\\
 &=&  \frac{2 r_0 \, \sin\phi_{0}\, \sin\theta_W }
             {1 + 2 r_0 \, \cos\phi_{0} \, \cos\theta_{W}  + r_0^2}
  \,,
\\[0.2em]
\eta_{\rm CP} \, S_{J/\psi K_S} &=& - \sin(2\beta)
  + \frac{2 r_0 \, \sin\theta_{W} \left( \cos(2\beta) \, \cos \phi_{0}
  + r_0 \, \cos(2\beta -\theta_{W}) \right)}
             {1 + 2 r_0 \, \cos\phi_{0} \, \cos\theta_{W}  + r_0^2} \,.
\end{eqnarray}  
We expect the NP amplitudes to provide small corrections to the SM,
$0 \leq r_0 \ll 1$, and thus to first approximation we have
\begin{eqnarray}
C_{J/\psi K_S} &\simeq &  2 r_0 \, \sin\theta_{W} \, \sin\phi_{0} \,,
\nonumber \\[0.2em]
\eta_{\rm CP} \, S_{J/\psi K_S} + \sin(2\beta) & \simeq & 
2 r_0 \, \sin\theta_W \,\cos\phi_{0} \,\cos(2\beta)
\,.
\label{approx}
\end{eqnarray}  
From this we read off the interesting parameter combinations
\begin{equation}
|r_0 \sin \theta_W| \simeq 
\frac{\sqrt{(\eta_{\rm CP} \, S_{J/\psi K_S}+\sin2\beta)^2 + (C_{J/\psi K_S} \, \cos2\beta)^2}}{2 \cos2\beta} \,,
\end{equation}
determining the {\em overall}\/ size of the deviations of $C$ from 0, 
and of $S$ from $\sin2\beta$, and
\begin{equation}
\tan \phi_{0} \simeq \frac{C_{J/\psi K_S} \, \cos2\beta}{\eta_{\rm CP} \, S_{J/\psi K_S} + \sin2\beta} \,,
\end{equation}
determining the {\em relative}\/ size of the two effects.

Notice that from the CP asymmetries alone we cannot
draw any conclusion about the value of the NP phase $\theta_W$.
This is a consequence of a reparameterization invariance 
(see e.g.\ \cite{London:2000aq}) which
leaves the decay amplitudes for the neutral $B$ decays in (\ref{NewAmp}),
as well as the branching fraction and the direct
CP asymmetry for the charged
$B$ decays invariant,
\begin{eqnarray}
&& {\cal A}_0' 
= {\cal A}_0 \left(1 + \xi \, (r_0 \, e^{i \phi_{0}}+r_1 \, e^{i\phi_1})\right) \,,
\nonumber \\[0.2em]
&& 
\cos\theta_W' = \frac{\cos\theta_W -\xi}{\sqrt{ 1-2 \, \xi \,\cos\theta_W + \xi^2}}
\,,
\qquad
\sin\theta_W' = \frac{\sin\theta_W}{\sqrt{ 1-2 \, \xi \,\cos\theta_W + \xi^2}}
\,,
\label{repar}
\end{eqnarray}
and similar transformations for the amplitude parameters $r_{0,1}$ and $\phi_{0,1}$,
where the parameter $\xi$ (and therefore also the values for $r_{0,1}$, 
$\phi_{0,1}$ and $\theta_W$) is arbitrary 
as long as the hadronic matrix element ${\cal A}_0$
for the leading SM contribution is not given explicitly.
In particular, for  $r_1=0$, $r_0 \ll 1$ and
small reparameterizations $\xi \ll 1$, we 
approximately have
\begin{eqnarray}
  r_0 &\to& r_0 \left( 1 - \xi \,\cos\theta_W + {\cal O}(\xi^2) \right) \,,
\qquad \phi_0 \to \phi_0 \left(1+ {\cal O}(\xi^2) \right) \,,
\cr  
  \sin\theta_W &\to& \sin\theta_W \left ( 1 + \xi \, \cos\theta_W + {\cal O}(\xi^2)
  \right) \,,
\label{reparapprox}
\end{eqnarray}
which explicitly shows the reparameterization invariance of
(\ref{approx}).
\begin{figure}[t!!!t]
\begin{center}
\parbox{0.4\textwidth}{
\rotatebox{90}{\hspace{0.1\textwidth}\footnotesize $r_0 \, \sin\theta_W$}
\includegraphics[width=0.36\textwidth]{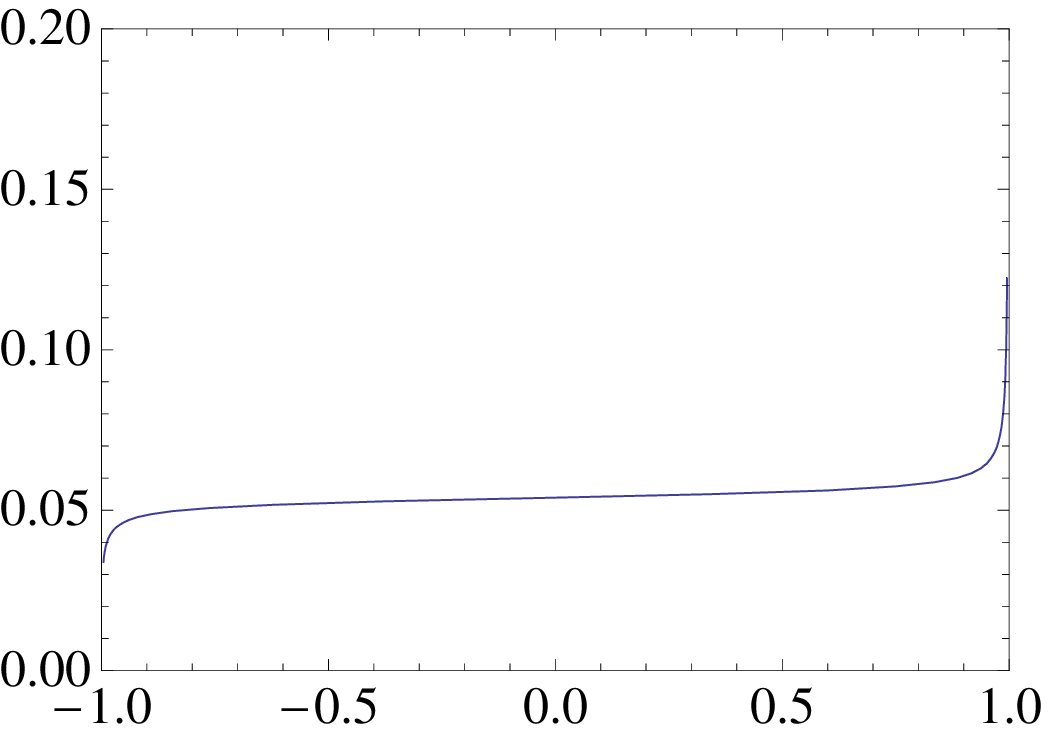}\\
\centerline{\footnotesize \ $\cos\theta_W$}}
\qquad
\parbox{0.4\textwidth}{
\rotatebox{90}{\hspace{0.11\textwidth}\footnotesize $\tan\phi_0$}
\includegraphics[width=0.36\textwidth]{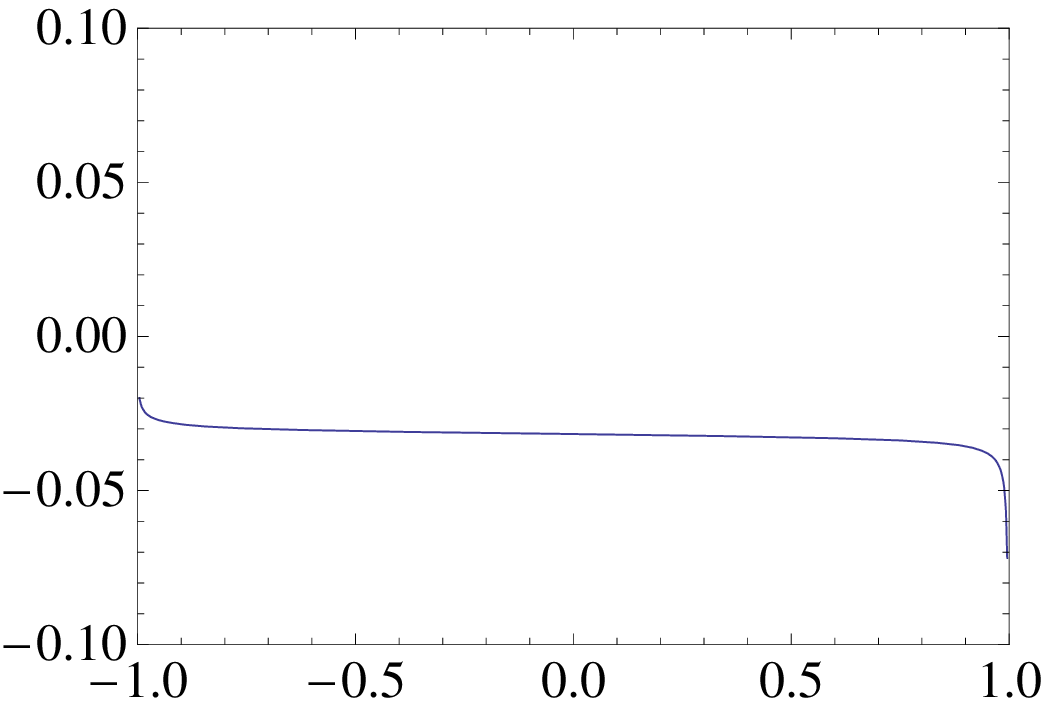}\\
\centerline{\footnotesize \\  $\cos\theta_W$
}}
\vspace{0.8em}
\parbox{0.9\textwidth}{\caption{\label{fig:repar} \small
  Illustration of the reparameterization invariance:
  The result for  $r_0 \, \sin\theta_W$ (left) and
  $\tan\phi_0$ (right) for the fit to $J/\psi K$ observables
  with NP contributions to $\Delta I=0$ as a function of $\cos\theta_W$.
  (The case of a SM-like NP phase is given by the central values
    $\cos\theta_W=-0.38$, $r_0 \, \sin\theta_W = 0.053$, $\tan\phi_0=-0.03$,
    corresponding to the fit in the last row of Table~\ref{tab:tab1} below.)
}}
\end{center}
\end{figure}
The reparameterization invariance is illustrated in Fig.~\ref{fig:repar},
where as an example we consider the fit result for a $\Delta I=0$
NP contribution with $\theta_W = \pi-\gamma_{\rm SM}$ found in the last row of Table~\ref{tab:tab1}
below, and apply the reparameterizations in (\ref{repar}) to generate
the equivalent solutions for other values of $\theta_W$. In particular,
we verify that the combinations $r_0 \, \sin\theta_W$ and $\tan \phi_0$
are approximately reparameterization-invariant, except for $\theta_W$
near zero or $\pi$.

As a consequence of the reparameterization invariance, the fit to the experimental data will generally allow for ''unphysical'' solutions, 
where the strong and weak phases are tuned in such a way
that the absolute size of the NP contribution $r_0$ can be unreasonably large.
In order to suppress such effects, we implement additional constraints:
(i) For small NP contributions, the fit should not depend on the parameter combination
$|r_0 \, \cos\theta_W|$; constraining $|r_0 \, \cos\theta_W|<0.4$ should therefore
only affect the unphysical solutions. (ii) If the phase $\theta_W$ of the NP operator
is close to the SM one, we do not expect to be sensitive to NP
in CP asymmetries in any case; we may therefore concentrate on
$30^\circ \leq \theta_W \leq 150^\circ$. (iii) For $\theta_W=\pi-\gamma_{\rm SM}$
our fit can also be interpreted as a determination of the size of sub-leading
SM contributions from Cabibbo- and penguin-suppressed amplitudes, which possibly
may have been underestimated in \cite{Boos:2004xp,Li:2006vq}.
In this case, one could also include the information from
$B \to J/\psi \pi$ decays to further constrain the hadronic parameters, 
using $SU(3)$ flavour symmetry \cite{Ciuchini:2005mg}, and correcting 
for the different relative CKM factors.
Considering the CP asymmetries in $B \to J/\psi \pi$ alone, we find that
the constraints on $r_0 \, e^{i\delta_0}$ 
are less restrictive than and consistent with the $B \to J/\psi K$ case.
The ratio of branching fractions in $B \to J/\psi \pi$ and 
$B \to J/\psi K$ further constrains $r_0$ \cite{Ciuchini:2005mg}.
However, we find that this ratio essentially depends on the 
combination
\begin{align}
\frac12 \, \frac{\Gamma[B^0 \to J/\psi K^0]}{\Gamma[B^0 \to J/\psi \pi^0]}
 & \approx   \frac{\lambda^2}{ R_{SU(3)}^2 \lambda^4 + r_0^2} \,,
\end{align}
and thus the constraints on $r_0$ are highly correlated with
the assumptions on the $SU(3)$ breaking parameter $R_{SU(3)}$
for the ratio of the leading $B \to J/\psi\pi$ and $B\to J/\psi K$
amplitudes. As this ratio cannot be estimated in a model-independent
way at present, we refrain from a detailed quantitative analysis.
However, it should be mentioned that for $R_{SU(3)}\approx 1$, 
smaller values for $r_0$ are favoured.

\begin{figure}[t!!!pbt!]
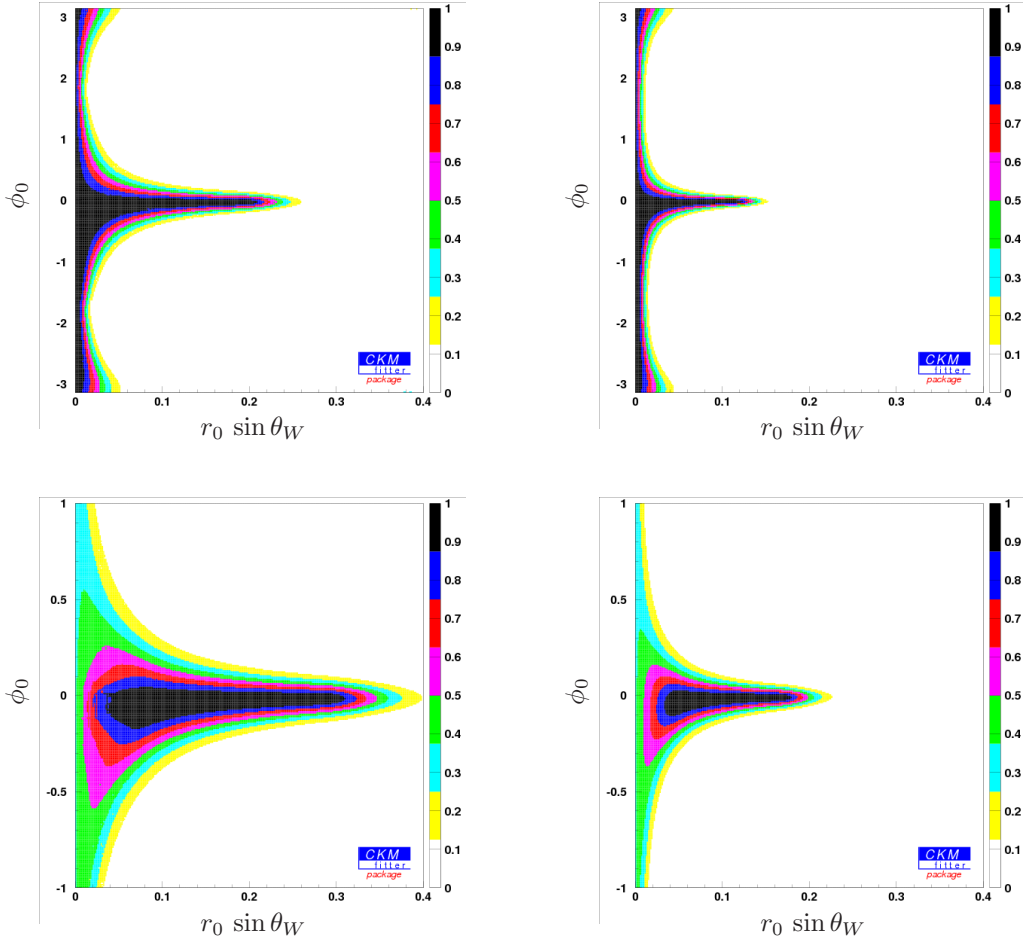

\begin{center}
\parbox{0.4\textwidth}{
\rotatebox{90}{\hspace{0.18\textwidth}\footnotesize $\phi_0$}
\includegraphics[width=0.35\textwidth]{1107_rswphis_free_rc.eps2}\\[-1em]
\centerline{\footnotesize \ $r_0 \, \sin\theta_W$}}
\qquad
\parbox{0.4\textwidth}{
\rotatebox{90}{\hspace{0.18\textwidth}\footnotesize $\phi_0$}
\includegraphics[width=0.35\textwidth]{1107_rswphis_fix.eps2}
\\[-1em]
\centerline{\footnotesize \ $r_0 \, \sin\theta_W$}}
\\[1.8em]
\parbox{0.4\textwidth}{
\rotatebox{90}{\hspace{0.18\textwidth}\footnotesize $\phi_0$}
\includegraphics[width=0.35\textwidth]{1107_rswphis_incl_free_rc_zoom.eps2} \\[-1em]
\centerline{\footnotesize \ $r_0 \, \sin\theta_W$}}
\qquad
\parbox{0.4\textwidth}{
\rotatebox{90}{\hspace{0.18\textwidth}\footnotesize $\phi_0$}
\includegraphics[width=0.35\textwidth]{1107_rswphis_fix_zoom.eps2}
\\[-1em]
\centerline{\footnotesize \ $r_0 \, \sin\theta_W$}}
\vspace{1.8em}
\parbox{0.9\textwidth}{ 
\caption{\small Fit results $\phi_0$ vs.\ $r_0 \, \sin\theta_W$
for different scenarios, see also Table~\ref{tab:tab1}.
The plots on the upper half refer to the case where $|V_{ub}|$ is determined
from exclusive and inclusive decays, whereas for the plots in the lower
half only the inclusive value is used. In the plots on the left only
the constraint $|r_0 \cos \theta_W|\leq 0.4$ is imposed. The plots on the
right are for fixed values $\theta_W= \pi-\gamma_{\rm SM}$. \label{fig:plot1}
}}
\end{center}
\end{figure}

Using the experimental values for $C_{J/\psi K_S}$, $S_{J/\psi K_S}$,
and $A_{\rm CP}^{\rm dir}(B^- \to J/\psi K^-)$,
together with the value for $\sin2\beta$ from the indirect
determination in Fig.~\ref{figNNPs2bga}, we fit the preferred ranges
for the NP parameters -- applying the different constraints
as discussed above -- as shown in Table~\ref{tab:tab1}
and Fig.~\ref{fig:plot1}. 
\begin{table}[t!!!t!]
\begin{center}
\parbox{0.9\textwidth}{
\caption{\label{tab:tab1} \small
Fit to direct and mixing-induced CP asymmetries in $B \to J/\psi K$,
using the indirect determination of $\sin2\beta$ and including 
the $\Delta I=0$ NP contribution $r_0$, only. We show the $1\sigma$ confidence level for
the two relevant parameter combinations $|r_0 \sin \theta_W|$ and $\phi_{0}$,
using different additional constraints to suppress ''unphysical'' solutions (see text). The 
upper half of the table corresponds to using the $\sin2\beta$ value from the indirect 
fit with $|V_{ub}|_{\rm excl.+incl.}$ in Fig.~\ref{figNNPs2bga}. In the lower half, only
$|V_{ub}|_{\rm incl.}$ from \cite{Lacker:2007me} is used.
}\vspace{0.8em}
}
\parbox[c]{0.47\textwidth}
{
}\qquad
\begin{tabular}{|lll|c|c|}
\hline
\multicolumn{3}{|l|}{Scenario} &  $|r_0 \, \sin \theta_W|$  &  $ \tan \phi_{0}$ \\ 
\hline \hline
excl.+incl. 
 & $\theta_W$ free & $|r_0 \cos \theta_W|\leq 0.4$
                                  & [0 to 0.23] & unconstrained \\
 & $30^\circ \leq \theta_W \leq 150^\circ$ & $|r_0 \cos \theta_W|$ free
                                  & [0 to 0.19] & unconstrained \\
 & $30^\circ \leq \theta_W \leq 150^\circ$ & $|r_0 \cos \theta_W|\leq 0.4$
                                  & [0 to 0.19] & unconstrained \\
 & $\theta_W= \pi-\gamma_{\rm SM}$  &  $|r_0 \cos \theta_W|$ free
                             &    [0 to 0.13] & unconstrained\\
\hline
incl. 
 & $\theta_W$ free & $|r_0 \cos \theta_W|\leq 0.4$
                                  & [0.02 to 0.34] & [-0.41 to 0.18] \\
 & $30^\circ \leq \theta_W \leq 150^\circ$ & $|r_0 \cos \theta_W|$ free
                                  & [0.03 to 0.33] & [-0.26 to 0.12] \\
 & $30^\circ \leq \theta_{W} \leq 150^\circ$ & $|r_0 \cos \theta_{W}|\leq 0.4$
                                  & [0.03 to 0.33] & [-0.26 to 0.12] \\
 & $\theta_W= \pi-\gamma_{\rm SM}$  &  $|r_0 \cos \theta_W|$ free
                             & [0.03 to 0.19] & [-0.24 to 0.11] \\
\hline
\end{tabular}
\end{center}
\end{table}
Since the value for $|V_{ub}|$ from the average of inclusive and exclusive
decays is close to its indirect determination from $\sin2\beta$,
the fitted range for $r_0 \, \sin\theta_W$ in this case is consistent with zero,
and the related strong phase $\phi_0$ is unconstrained. Still, for sufficiently
small strong phases, NP contributions of the order 20\%\ are not
excluded either.  On the other hand, taking into account the 
inclusive value of $|V_{ub}|$ (with its
small tension with $\sin2\beta$) only,
the fit prefers non-zero values for $r_0 \, \sin\theta_W$ of 
the order 5-30\%\ and relatively small strong phases $\phi_0$.
(Notice that small strong phases are generally expected within 
 the QCD factorization approach to hadronic matrix elements
 in the heavy-quark limit \cite{Beneke:2000ry}.)
Compared to the estimate of SM corrections in
\cite{Boos:2004xp,Li:2006vq}, the typical order of magnitude for $r_0$ is thus
significantly larger.
Although the present experimental situation is not conclusive,
our analysis shows that an improvement of the experimental 
precision for $B \to J/\psi K$ observables or
of the theoretical precision in the $|V_{ub}/V_{cb}|$ determination
may still lead to interesting conclusions.

\subsubsection{Fit with $\Delta I=0,1$ 
(new physics in $b \to s u \bar u$ or $b \to s d\bar d$)}

\begin{figure}[t!!bt]
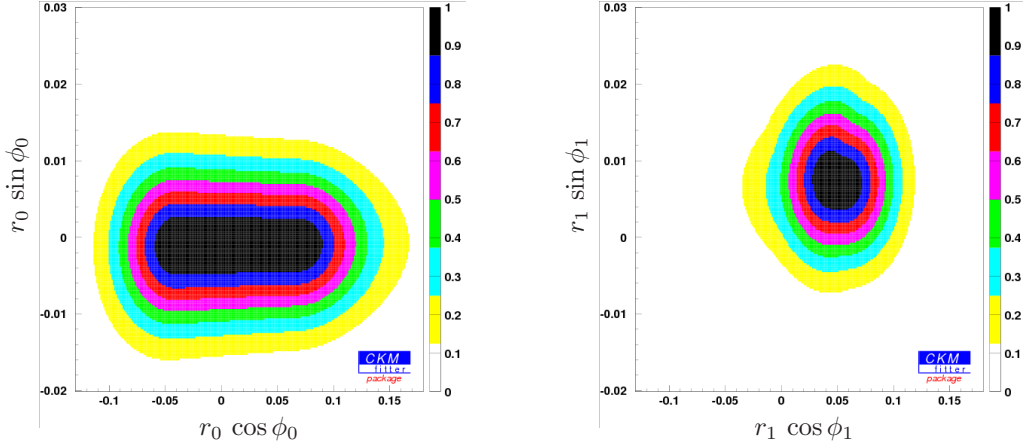

\begin{center}
\parbox{0.4\textwidth}{
\rotatebox{90}{\hspace{0.16\textwidth}\footnotesize $r_0 \, \sin\phi_0$}
\includegraphics[width=0.35\textwidth]{0907_r0_fix.eps2}\\[-1em]
\centerline{\footnotesize   $r_0 \, \cos\phi_0$}}
\qquad
\parbox{0.4\textwidth}{
\rotatebox{90}{\hspace{0.16\textwidth}\footnotesize $r_1 \, \sin\phi_1$}
\includegraphics[width=0.35\textwidth]{0907_r1_fix_unzoom.eps2}\\[-1em]
\centerline{\footnotesize  $r_1 \, \cos\phi_1$
}}

\vspace{0.8em}
\parbox{0.9\textwidth}{\caption{\label{fig:jpsi_i1} \small
  The result for  $r_0 \, e^{i\phi_0}$ (left) and
  $r_1 \, e^{i \phi_1}$ (right) in the complex plane
  from the fit to $J/\psi K$ observables, with isospin-breaking 
  NP contributions $b \to s u\bar u$ or $b \to s d\bar d$. 
  The new weak phase has been
  fixed to $\phi_W = \pi-\gamma_{\rm SM}$.
}
}
\end{center}
\end{figure}

New physics contributions to either $b \to s u\bar u$ or $b \to s d\bar d$
may lead to isospin asymmetries between charged and
neutral $B \to J/\psi K$ decay rates and CP asymmetries. In this
case we may fit (\ref{NewAmp}) with both $r_0\neq 0$ and $r_1 \neq 0$,
and consider the observables in Fig.~\ref{fig:plot1}
together with the (CP-averaged) isospin breaking in the decay 
rates \cite{Yao:2006px}
\begin{eqnarray}
A_I(B \to J/\psi K) = \frac{
  \Gamma[B_d \to J/\psi K_0] -\Gamma[B^\pm\to J/\psi K^\pm]}{
  \Gamma[B_d \to J/\psi K_0] +\Gamma[B^\pm\to J/\psi K^\pm]}
&=& 
  -0.035 \pm 0.026 \,.
\end{eqnarray}
For small values of $r_0$ and $r_1$, following \cite{Fleischer:2001cw}, 
we have the approximate relations
\begin{eqnarray}
\eta_{\rm CP} \, S + \sin2\beta &\simeq& \phantom{-} 2 \left( r_0 \, \cos\phi_{0} 
  + r_1 \, \cos\phi_1 \right)  \sin\theta_W
\, \cos2\beta \,,
\cr
  A_{\rm CP}^{\rm avg} &\simeq& -2 \, r_0 \, \sin\phi_{0} \, \sin\theta_W  \,,\cr
  \Delta A_{\rm CP} &\simeq& -2 \, r_1 \, \sin\phi_{1} \, \sin\theta_W  \,,\cr
  A_I 
  &\simeq& \phantom{-} 2 \, r_1 \, \cos\phi_{1} \, \cos\theta_W \,.
\label{approx2}
\end{eqnarray}
They are manifestly invariant
under the approximate reparameterizations,
following from (\ref{repar}) in the limit $\xi = {\cal O}(r_{0,1}) \ll 1$,
\begin{eqnarray}
\sin\theta_W &\to& \sin \theta_W \left(1 + \xi \, \cos \theta_W + {\cal O}(\xi^2)
\right) \,, 
\cr 
 \cos\theta_W &\to& \cos \theta_W - \xi \, \sin^2 \theta_W + {\cal O}(\xi^2) \,, 
\cr
  r_0 \, \cos\phi_0 
+ r_1 \, \cos\phi_1
& \to & 
(r_0 \, \cos\phi_0 + r_1 \, \cos\phi_1) 
\left(1 - \xi  \, \cos\theta_W 
  +{\cal O}(\xi^2) \right)\,, \cr
 r_1 \, \cos \phi_1 & \to & r_1\, \cos\phi_1 \left(1  + \xi \, \sin\theta_W \, \tan\theta_W + {\cal O}(\xi^2) \right) \,,
\cr 
r_{0,1} \, \sin\phi_{0,1}& \to & r_{0,1} \, \sin\phi_{0,1} \left( 1 - \xi  \, \cos\theta_W 
  +{\cal O}(\xi^2) \right)  
\,.
\label{reparapprox2}
\end{eqnarray}
To keep the discussion simple, we may again concentrate on the special
case $\theta_W = \pi -\gamma$. The fit result is plotted in Fig.~\ref{fig:jpsi_i1}.
The $1\sigma$ parameter ranges are given by
\begin{eqnarray}
&& r_0 \, \cos\phi_0 = \left[-0.077 \ {\rm to} \ 0.112 \right] \,, \qquad
   r_0 \, \sin\phi_0 = \left[-0.008 \ {\rm to} \ 0.006 \right] \,,
\cr
&&  r_1 \, \cos\phi_1 = \left[\phantom{-}0.013  \ {\rm to} \ 0.088 \right] \,, \qquad
  r_1 \, \sin\phi_1 = \left[\phantom{-}0.000 \ {\rm to} \ 0.015 \right] \,.
\label{fit:fit2}
\end{eqnarray}
Notice that again, the strong phases for the preferred ranges
turn out to be small.
Solutions for other values of $\theta_W$ can
be reconstructed by means of the reparameterization invariance 
(\ref{reparapprox2}).

We conclude that small deviations from the SM expectations
in $B \to J/\psi K$ can be explained by NP in
either $b \to s u\bar u$ or $b \to s d\bar d$, alone. However,
one has to keep in mind that, compared to the contributions from
$b \to s c\bar c$, the
$b \to s u\bar u$ or $b \to s d\bar d$ only contribute via
penguin ($r_0$) or annihilation ($r_1$) diagrams to
hadronic matrix elements. 
Thus, an additional suppression with respect to the tree-level
matrix elements fitted in the last section (see Table~\ref{tab:tab1}) 
is expected.
Notice that, depending on the actual size of these suppression factors, our 
result for $r_0$ and $r_1$ may also be interpreted as due to 
unexpectedly large effects from sub-leading SM operators.
Again, the information from $B \to J/\psi \pi$ observables together
with assumptions on $SU(3)$ breaking effects could be
used to further constrain $r_0$ and $r_1$ in this case.

\subsection{Analysis of $B \to \phi K$}

The discussion of $B \to \phi K$ decays is very similar to the $B \to J/\psi K$
case. The most important difference is due to the fact that a tree-level 
operator for $b \to ss\bar s$ transitions is absent in the SM, and therefore
the leading SM amplitude ${\cal A}_0(B \to \phi K)$ already
receives a penguin
suppression factor of order $\lambda$ compared to ${\cal A}_0(B \to J/\psi K)$ 
(see for instance \cite{Fleischer:2001pc}). Consequently, the relative
size of both, Cabibbo suppressed SM contributions as well as
potential NP contributions, may be enhanced accordingly.
Indeed, the experimentally observed
discrepancy between $S_{\phi K_S}$ and $\sin2\beta$ is more pronounced,
while estimates within the SM typically give small effects 
\cite{Grossman:2003qp,Williamson:2006hb,Gronau:2004hp,Cheng:2005bg,Beneke:2005pu}.

To keep the notation simple, 
we use the same symbols $r_i$, $\phi_{i}$ as in the $B \to J/\psi K$ 
to parameterize NP contributions to the $B \to \phi K$ decay
amplitudes
\begin{eqnarray} 
{\cal A} (\bar B\to \phi \bar K) 
 &=& {\cal A}_0 (\bar B \to \phi K) 
 \left[ 1 + r_0 \, e^{i \theta_W}  e^{i \phi_{0}} \mp
    r_1 \, e^{i \theta_W}  e^{i \phi_{1}} \right] \,.
\label{NewAmpPhi}
\end{eqnarray}
However, one has to keep in mind that both,
the involved NP operators and the strong dynamics in hadronic
matrix elements, are different.

\subsubsection{Fit with $\Delta I=0$ (new physics in $b \to s s \bar s$)}

\begin{figure}[t!bt]
\begin{center}
\parbox[c]{0.48\textwidth}{
\rotatebox{90}{\hspace{0.21\textwidth}\footnotesize $\phi_0$}
\includegraphics[width=0.42\textwidth]{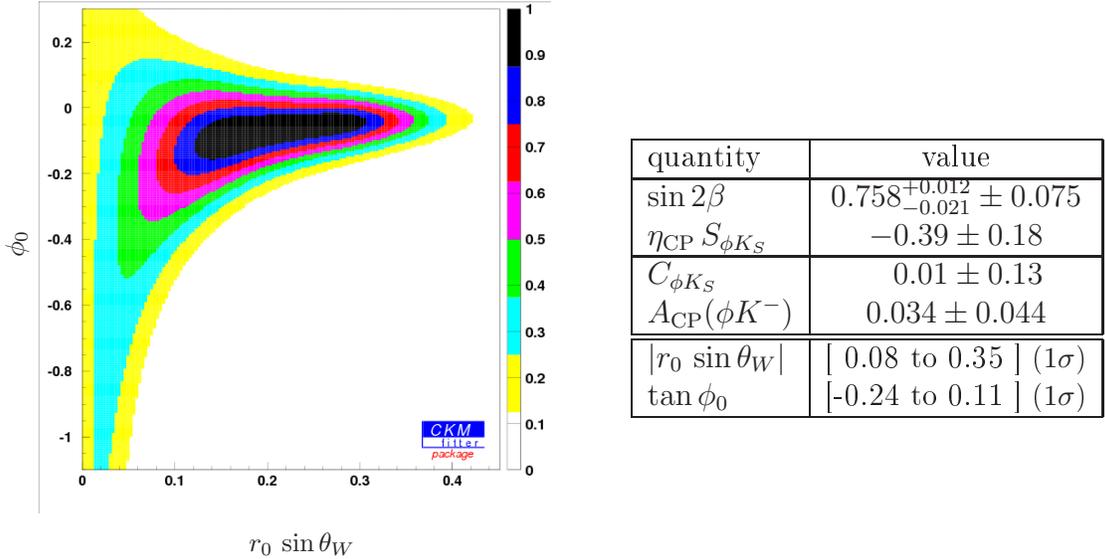}
\\
\centerline{\footnotesize \ $r_0 \, \sin\theta_W$}}
\quad
{\begin{tabular}{|l|c|}\hline
 quantity & value\\\hline
 $\sin2\beta$ & $0.758^{+0.012}_{-0.021} \pm 0.075$\\
 $\eta_{\rm CP}\,S_{\phi K_S}$ & $ -0.39 \pm 0.18$ \\ \hline
 $C_{\phi K_S}$ & $ \phantom{+}0.01 \pm 0.13$ \\
 $A_{\rm CP}(\phi K^-)$ & $0.034 \pm 0.044 $\\
\hline \hline
 $|r_0 \, \sin \theta_W|$ & [\phantom{-}0.08 to 0.35 ] {\small ($1\sigma$)}\\
 $ \tan \phi_{0}$ & [-0.24 to 0.11 ] {\small ($1\sigma$)}\\
\hline
\end{tabular}}
\vspace{0.8em}
\parbox{0.9\textwidth}{\caption{\label{fig:plot2} \small
Fit to direct and mixing-induced CP asymmetries in $B \to \phi K$,
using the indirect determination of $\sin2\beta$ and including 
the contribution of a NP operator with $\Delta I=0$, i.e.\ $b \to s s\bar s$.
The NP weak phase is set to $\theta_W = \pi -\gamma_{\rm SM}$.
Left: Confidence levels for
the two relevant parameter combinations $|r_0 \sin \theta_{W}|$ and $\phi_{0}$.
Right: Input parameters (upper half \cite{Barberio:2006bi}) and 
$1\sigma$~ranges for the output values (lower half) of the fit.
}}
\end{center}
\end{figure}

Using the experimental values for the direct and mixing-induced
CP asymmetries in $B \to \phi K$
together with the value for $\sin2\beta$ from the indirect
determination in Fig.~\ref{figNNPs2bga}, we fit the preferred ranges
for the NP parameters as shown in Fig.~\ref{fig:plot2}. 
Again, we only quote the result for a particular value 
for the new weak phase, $\theta_W=\pi -\gamma_{\rm SM}$. Other solutions
follow from the same reparameterization invariance as in (\ref{repar}).
Comparison with the $B \to J/\psi K$ case in Fig.~\ref{fig:plot1} shows:
\begin{itemize}
 \item Again, the fit prefers small strong phases $\phi_0$.

 \item The preferred value for $r_0$ in $B \to \phi K$ is by a factor of
       2-3 larger than the one in $B \to J/\psi K$.
       After correcting for the penguin suppression factor, phase space and
       normalization, this implies
       that the coefficients of the involved NP operators in both cases
       may be of similar size. 
\end{itemize}
We emphasize, that the latter observation also implies
that unusually large hadronic penguin matrix elements in
the SM could simultaneously explain the $B \to J/\psi K$ 
and $B \to \phi K$ discrepancies.

\subsubsection{Including $\Delta I=1$ operators}

The current data shows no evidence for 
isospin asymmetries in $B \to \phi K$ decays 
\cite{Barberio:2006bi},
\begin{eqnarray}
\Delta A_{\rm CP}(B \to \phi K) &=& 0.02 \pm 0.13  \,,\cr
 A_I(B \to \phi K) &=&  0.04 \pm 0.08 \,,
\end{eqnarray}
although again the \emph{relative} effects from
$b\to s u\bar u$ and $b \to s d\bar d$ operators are expected
to be larger than in the $B \to J/\psi K$ case.
We find it instructive to turn the argument around and
estimate the potential size of isospin violation in 
$B \to \phi K$ by simply rescaling the solutions for $r_0$
and $r_1$ in (\ref{fit:fit2}) by a factor $2.5$ (see above), which yields
the ``1-$\sigma$ estimates''
\begin{eqnarray}
\Delta A_{\rm CP}(B \to \phi K) 
&\stackrel{?}{\sim}& \phantom{-} (0 \mbox{\ to\ } 0.14) \,,\cr
  A_I (B \to \phi K)&\stackrel{?}{\sim}& - (0.17 \mbox{\ to\ } 0.01) \,.
\end{eqnarray}
The resulting order of magnitude is comparable with
the present experimental uncertainties. If our estimate makes sense,
a moderate improvement of the experimental sensitivity could already 
lead to a positive signal for isospin violation in $B \to \phi K$.


\subsection{Analysis of $B \to K \pi$}

In the SM, the general isospin decomposition for $B \to K\pi$ decays can
be parameterized as \cite{Neubert:1998re,Beneke:2001ev}
\begin{eqnarray}
  {\cal A}(B^- \to \pi^- \bar K^0) &=&
    {P} \left( 1 + {\epsilon_a \, e^{i\phi_a}} \, e^{-i\gamma} \right) \,,
\nonumber \\
 -\sqrt2\, {\cal A}(B^- \to \pi^0  K^-) &=&
    {P} \left( 1 + {\epsilon_a \, e^{i\phi_a}} \, e^{-i\gamma} 
     - {\epsilon_{3/2} \, e^{i \phi_{3/2}}} \left( e^{-i\gamma}- 
     { q e^{i\omega}} \right)
\right)
\,, \nonumber \\
 - {\cal A}(\bar B_d \to \pi^+  K^-) &=&
   {P} \left( 1 + {\epsilon_a \, e^{i\phi_a}} \, e^{-i\gamma}
   - {\epsilon_{T} \, e^{i \phi_T}} \left( e^{-i\gamma}- 
     {q_C e^{i\omega_C}}
\right)\right)
\label{piKSM}
\end{eqnarray}
and 
$$
\sqrt2 \, {\cal A}(\bar B_d \to \pi^0 \bar K^0)= 
{\cal A}(B^-\to \pi^- \bar K^0)
+ \sqrt2 \, {\cal A}(B^- \to \pi^0 K^-)
- {\cal A}(\bar B_d \to \pi^+ K^-)
$$ 
fixed by isospin symmetry (i.e.\ neglecting QED
and light quark-mass corrections in the hadronic matrix elements).
Here $P$ is the dominating penguin amplitude, whereas the quantities $\epsilon_{T,3/2}$
contain tree-operators but are doubly CKM-suppressed. 
Without any assumptions on strong interaction dynamics, in the isospin
limit one is left with 11 independent hadronic parameters
for 9 observables.
In order to test the SM against
possible NP effects in these decays, one needs additional dynamical input.
Qualitative results from QCDF \cite{Beneke:2001ev} include:
\begin{itemize}
   \item The $SU(3)_F$ symmetry prediction \cite{Fleischer:1995cg}
\begin{equation}
q \, e^{i\omega} \simeq - \frac{3}{2} \, \frac{|V_{cb} V_{cs}^*|}{|V_{ub} V_{us}^*|} \,
    \frac{C_9+C_{10}}{C_1+C_2} 
\label{qrel}
\end{equation}
 only receives small corrections.

   \item The parameter $\epsilon_a \, e^{i\phi_a}$ is negligible in QCDF.
      Consequently the direct CP asymmetry in $B^- \to \pi^- K^0$ is tiny
      (in accord with experiment).

   \item The parameter $q_C \, e^{i \omega_C}$ is of minor numerical importance.

   \item The parameters $\epsilon_T$ and $\epsilon_{3/2}$ are expected to be
         of the order 20-30\%, with the related strong phases of the order  $10^\circ$. Furthermore, at least at NLO accuracy, the difference between 
        $\epsilon_T \, e^{i \phi_T}$ and $\epsilon_{3/2} \, e^{i \phi_{3/2}}$ is
        a sub-leading effect proportional to the small coefficients $a_{2,7,9}$
        in QCDF.

\end{itemize}
In the subsequent fits, we will set $\epsilon_a$
to zero and use the values from \cite{Beneke:2001ev},
\begin{eqnarray}
&& q = 0.59 \pm 0.12 \pm 0.07 \,,\qquad
 \omega = -0.044 \pm 0.049 \,, 
\label{q}
\\
&& q_C = 0.083 \pm 0.017 \pm 0.045 \,,\quad
 \omega_C = - 1.05 \pm 0.86 \,,
\label{qc}
\end{eqnarray}
in order to reduce the number of independent hadronic parameters within
the SM to 5. (Notice that the overall penguin amplitude parameter $P$
in (\ref{piKSM}) will not be constrained from theory, but will essentially
be fixed by the experimental data for the $B^\pm \to \pi^\pm K^0$
branching fractions.)
Tensions in the fit, or incompatible values for the parameters
$\epsilon_{T,3/2}$ and $\phi_{T,3/2}$ then may be taken as indication for possible  
NP contributions. 

\subsubsection{New physics in $B \to K\pi$ ?}

\label{sec:npkpi}

The critical observables in  $B \to K\pi$ transitions
are  \cite{Fleischer:2007mq}
\begin{eqnarray}
 R_c &=& 2 \left[ 
  \frac{{\rm BR}(B^- \to \pi^0 K^-) + {\rm BR}(B^+ \to \pi^0 K^+)}
{{\rm BR}(B^- \to \pi^- \bar K^0) + {\rm BR}(B^+ \to \pi^+ 
 K^0)}\right]
= 1.11 \pm 0.07 \,,
\nonumber
\\[0.2em]
 R_n &=& \frac12 \left[ 
  \frac{{\rm BR}(\bar B_d \to \pi^+ K^-) + {\rm BR}(B_d \to \pi^- K^+)}
{{\rm BR}(\bar B_d \to \pi^0 \bar K^0) + {\rm BR}(B_d \to \pi^0  K^0)}\right]
= 0.97 \pm 0.07 \,,
\nonumber
\\[0.4em]
\Delta A &=& A_{\rm CP}^{\rm dir}(B^\pm \to \pi^0 K^\pm)
- A_{\rm CP}^{\rm dir}(B_d \to \pi^\mp K^\pm) = 0.142 \pm 0.029 \,,
\label{CritObs}
\nonumber
\\[0.4em]
C_{\pi^0 K_S} &=& 0.14 \pm 0.11 \,, \qquad \eta_{\rm CP} \, S_{\pi^0 K_S} \ = \ -0.38 \pm 0.19 \,.
\end{eqnarray}
Within our SM approximation, we expect (see also \cite{SR})
\begin{eqnarray}
R_c- R_n &\simeq& 2 \,\epsilon_{3/2} \left( \epsilon_T
- \epsilon_{3/2} \left( 1-q^2  \right) \right) +
       {\cal O}(\lambda^3) \,,\\[0.2em]
\Delta A \simeq C_{\pi^0 K_S} 
 &\simeq& 2 \left( \epsilon_{T} \, \sin \phi_{T} - \epsilon_{3/2} \, \sin\phi_{3/2} \right)
 + {\cal O}(\lambda^3)
\,,\\[0.2em]
\eta_{\rm CP} \, S_{\pi^0 K_S} 
 &\simeq& - \sin 2\beta  +  2 \, \cos2\beta 
 \left( \epsilon_{T}  - \epsilon_{3/2} \right)
+ {\cal O}(\lambda^2)\,,
\end{eqnarray}
where we used that $\epsilon_{T,3/2} \sim\lambda$,
$\phi_{T,3/2} \sim \lambda$, $q_c \simeq 0$, $\omega \simeq0$, and $\cos\gamma \sim \lambda$  in the SM.
Considering the recent experimental data, 
the first relation turns out to be well fulfilled, whereas
the second and third relation require a sizeable difference between $\epsilon_{T} \, e^{i \phi_{T}}$ and $\epsilon_{3/2} \, e^{i\phi_{3/2}}$.

To quantify this observation, we perform a fit (within the SM) 
to the quantities 
$\epsilon_T \, e^{i\phi_T}$ and $\epsilon_{3/2} \, e^{i\phi_{3/2}}$, 
as shown in 
Table~\ref{tab:epsT32}.
\begin{table}[t!bpt!]
\begin{center}
\parbox{0.9\textwidth}{\caption{\small
\label{tab:epsT32} SM fit results for 
$\epsilon_T$, $\phi_T$, $\epsilon_{3/2}$, $\phi_{3/2}$, 
with $\epsilon_a=0$ and $q \, e^{i\omega}$ and $q_C \, e^{i \omega_C}$
varied according to (\ref{q},\ref{qc}) from \cite{Beneke:2001ev}.
The best fit values for the latter parameter are obtained 
as $q=0.49$, $\omega=0.005$, $q_C=0.038$, $\omega_C=-1.91$.
For comparison, we show in the last line estimates for the corresponding
hadronic parameters from \cite{Fleischer:2007mq}
which have been obtained by relating
$B \to \pi K$ to $B \to \pi\pi$ via $SU(3)$ relations and dynamical assumptions
(central values only).}
\vspace{0.8em}}
  \begin{tabular}{|l||c |c |c |c|c|c|}
\hline                & $\epsilon_T$    & $\phi_T$             & $\epsilon_{3/2}$ & $\phi_{3/2}$         & ${\rm Re} \, \Delta\epsilon$ & ${\rm Im} \, \Delta\epsilon$\\
\hline
Best:
& 0.21           
& 0.21	              
& 0.04            
& 0.07               
& 0.18               
& 0.07      
\\
$1\sigma$:
& [0.10, 0.32]   
& [0.10,0.50]          
& [0.01,0.15]      
& [0.05,0.09]          
& [0.07,0.33]         
& [0.05,0.09]
\\
$2\sigma$:
& [0.05,0.44]   
& [0.05,1.32]        
& [0.00,0.38]      
& [0.03,0.11]          
& [-0.13,0.42]        
& [0.03,0.11]
\\
\hline
\hline                & $r$    & $\delta$             & $r_c$ & $\delta_c$        
 & $-\rho_n \, \cos\theta_n$ & $- \rho_n \, \sin\theta_n$\\
\hline
\cite{Fleischer:2007mq} & 
 $0.12$ & $0.44$ & $0.20$ & $0.02$ & $-0.10$ & $0.04$
\\
\hline
\end{tabular}
\end{center}
\end{table}
In Table~\ref{tab:bestfit} (3rd column)
we compare the best fit result with experimental
data and observe a very good agreement.
In particular, the expected
approximate equality $\Delta A \simeq C_{\pi^0 K_S}$
is fulfilled by the data. 
The fitted values for the individual amplitude parameters 
$\epsilon_T$, $\phi_T$, $\epsilon_{3/2}$, $\phi_{3/2}$ are in qualitative
agreement with the expectations from QCDF. However, the comparison
of $\epsilon_T \, e^{i\phi_T}$ and $\epsilon_{3/2} \, e^{i\phi_{3/2}}$ shows
sizeable deviations,
$$
  \Delta \epsilon :=  \epsilon_T \, e^{i \phi_T}- \epsilon_{3/2} \, e^{i \phi_{3/2}}
 \neq 0\,,
$$
which are incompatible with the NLO predictions from
QCDF (for the status of NNLO predictions, see 
\cite{Beneke:2005,Bell:2006,Kivel:2006xc,Pilipp:2007mg}).
In the notation for topological amplitudes \cite{Gronau:1995hn} 
this would correspond to\footnote{%
In a parameterization where the annihilation topology $\tilde A$ is explicit
\cite{Neubert:1998re},
one has $|\Delta \epsilon/\epsilon_T|= (\tilde C+\tilde A)/(\tilde T-\tilde A)$,
where $\tilde T(\tilde C)$ denote the colour-allowed(-suppressed) tree amplitude.
}
a ratio $C/T = |\Delta \epsilon/\epsilon_T|$
in the range $[0.52-3.00]$ with the central value at $0.89$.
Assuming that higher-order QCD effects and non-factorizable
power corrections cannot substantially change the approximate equality between
$\epsilon_T$ and $\epsilon_{3/2}$, this might be taken as a weak indication
of NP in $B \to K\pi$ decays (for a recent discussion, see also
\cite{Baek:2007yy}). It is also interesting to compare the fitted
values for $\Delta \epsilon$ with the latest estimates obtained
in \cite{Fleischer:2007mq} on the basis of $SU(3)$ relations and
dynamical assumptions about sub-leading decay topologies, 
see last row in Table~\ref{tab:epsT32}. In this case, a sizeable
$C/T$ ratio is obtained from a fit to the $B \to \pi\pi$ observables,
but with the ``wrong'' sign for the corresponding strong amplitude,
compared to our SM fit.
As the dynamical mechanism for generating (sizeable) strong phases
in charmless non-leptonic $B$ decays is not completely understood,
a resolution of the observed discrepancies in $\Delta \epsilon$
from non-factorizable QCD corrections within the SM 
cannot be excluded
(see, for instance, the discussion in \cite{Feldmann:2004mg}).

\begin{figure}[t!!pbt]
\begin{center}
\parbox{0.48\textwidth}{
\rotatebox{90}{\hspace{0.2\textwidth}\footnotesize ${\rm Im}(\Delta \epsilon)$}
\includegraphics[width=0.42\textwidth]{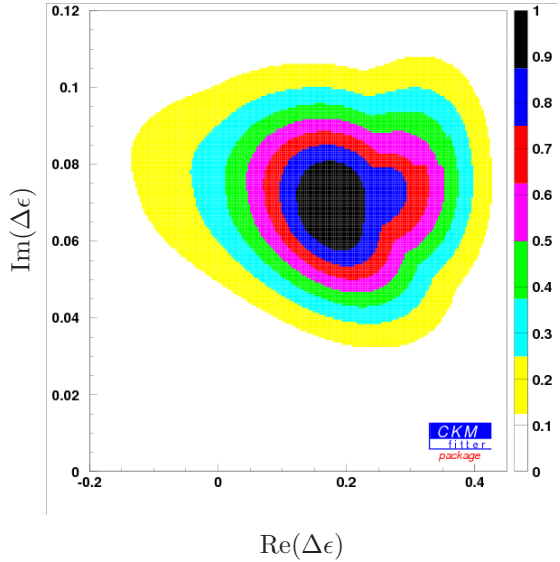}\\
\centerline{\footnotesize   ${\rm Re}(\Delta \epsilon)$}}
\vspace{0.8em}
\parbox{0.9\textwidth}{\caption{\small \label{fig:KpiSMdiff}
SM fit results for 
$\Delta \epsilon $
in the complex plane, with $\epsilon_a=0$ and $q \, e^{i\omega}$ and $q_C \, e^{i \omega_C}$
taken from \cite{Beneke:2001ev}.
}}
\end{center}
\end{figure}

  \begin{table}[t!bt!]
  \begin{center}
  \parbox{0.9\textwidth}{\caption{\small \label{tab:bestfit}
  Experimental data for $B\to K\pi$-decays vs.\ various best fit
  results.
  The third column shows the SM fit with $\Delta \epsilon \neq 0$,
  which corresponds to $\chi^2/{\rm d.o.f.}=2.43/3$.
  The fourth column shows the best fit
  result for $\Delta \epsilon = 0$
  (with $\epsilon_T \, e^{i\phi_T}=\epsilon_{3/2} \, e^{i\phi_{3/2}}$ varied
  according to their QCDF ranges, see text)
  and a NP contribution with $\Delta I=0$
  and $\theta_W=\pi-\gamma_{\rm SM}$, yielding
  $\chi^2/{\rm d.o.f.}=18.5/6$.
  The last column shows the analogous fit
  result for  a NP contribution from (essentially) $b \to s u\bar u$
  with $\theta_W = \pi-\gamma_{\rm SM}$,
  which corresponds to $\chi^2/{\rm d.o.f.}=2.91/3$.
  Experimental values taken from HFAG \cite{Barberio:2006bi}.}
  \vspace{0.8em}}
  \begin{tabular}{|c|c||c|c|c|}
  \hline Observable & HFAG (after ICHEP'06) & SM fit
  & NP $(I=0)$ & NP $(I=0,1)$
  \\
  \hline
  $\overline{BR}(\pi^0K^-) \cdot 10^{6}$            & $12.8 \pm 0.6$
  & $12.2$ & $12.6$ & $12.6$
  \\
  $\overline{BR}(\pi^-\bar{K}^0)\cdot 10^{6}$       & $23.1 \pm 1.0$  &
  $23.9$ & $23.8$ & $23.8$
  \\
  $\overline{BR}(\pi^+K^-)\cdot 10^{6}$       & $19.4 \pm 0.6$  &
  $19.7$  & $19.6$ & $19.6$
  \\
  $\overline{BR}(\pi^0\bar{K}^0)\cdot 10^{6}$ & $10.0 \pm 0.6$  &
  $\phantom{1}9.5$  & $\phantom{1}9.0$ & $\phantom{1}9.2$
  \\\hline
  $\mathcal{A}_{CP}(\pi^-\bar{K}^0)$          & $\phantom{-}0.009\pm0.025$ &
  $0^*$ & $-0.02\phantom{0}$ & $0^*$
  \\
  $\mathcal{A}_{CP}(\pi^0K^-)$                & $\phantom{-}0.047\pm0.026$ &
  $\phantom{-}0.048$ & $\phantom{-}0.001$ & $\phantom{-}0.049$
  \\
  $\mathcal{A}_{CP}(\pi^+K^-)$                & $-0.095\pm0.015$       &
  $-0.095$ & $-0.06\phantom{0}$ & $-0.094$
  \\\hline
  $ \eta_{\rm CP} \, S_{\pi^0 K_S}$                            & $-0.38\pm0.19$         &
  $-0.39$  & $-0.34$ & $-0.48$
  \\
  $C_{\pi^0K_S}$                            & $\phantom{-}0.12\pm0.11$
  & $\phantom{-}0.14$ & $\phantom{-}0.06$ & $\phantom{-}0.13$
  \\\hline \hline
  $R_c$                                       & $1.11 \pm 0.07$        &
  $1.02$ & 1.06 & $1.06$
  \\
  $R_n$                                       & $0.97 \pm 0.07$        &
  $1.04$ & 1.09 & $1.07$
  \\
  $\Delta A$                                  & $0.142 \pm 0.029$ &
  $0.143$ &
  0.06 & $ 0.143$
  \\\hline
  \end{tabular}
  \end{center}
  \end{table}

We may interpret the required difference between $\epsilon_T$ and $\epsilon_{3/2}$ as
due to NP contributions in the $\Delta I=1$ Hamiltonian. In this case the
fit result for the quantity
$
  \Delta \epsilon 
$, shown in Fig.~\ref{fig:KpiSMdiff},
is already a measure for the possible effect of NP operators. Notice however,
that again the weak phase associated with these operators cannot be fixed.
To continue, we follow a similar line as in the analysis of $B \to J/\psi K$ and 
$B \to \phi K$ decays, and assume that only one particular NP operator of the type 
$b \to s q\bar q$ gives a significant contribution in $B \to K\pi$ decays.


\subsubsection{New physics contributions with $\Delta I=0$ only}

The presence of a NP contribution with $\Delta I=0$ (in our case,
this includes the ''charm penguin'' $b \to s c\bar c$, as well as
$b\to s s\bar s$ and $b \to s (u\bar u + d\bar d)$) has
the same impact as the SM parameter $\epsilon_a$ in (\ref{piKSM}), except for a
possibly different weak phase.
Within our approximation, one thus obtains
\begin{eqnarray}
  {\cal A}(B^- \to \pi^- \bar K^0) &\simeq &
    {P} \left( 1 + r_0 \, e^{i\phi_0} \, e^{i \theta_W} \right) \,,
\nonumber \\
 -\sqrt2\, {\cal A}(B^- \to \pi^0  K^-) &\simeq &
    {P} \left( 1 + r_0 \, e^{i\phi_0} \, e^{i\theta_W} 
     - \epsilon_{3/2} \, e^{i \phi_{3/2}} \left( e^{-i\gamma} -
      q e^{i\omega}\right)
\right)
\,, \nonumber \\
 - {\cal A}(\bar B_d \to \pi^+  K^-) &\simeq&
   {P} \left( 1 + r_0 \, e^{i\phi_0} \, e^{i \theta_W}
   - \epsilon_{T} \, e^{i \phi_T} \left( e^{-i\gamma} -
      q_C \, e^{i\omega_C}\right)\right) \,,
\end{eqnarray}
where $r_0 \, e^{i\phi_0} \, e^{i \theta_W}$ parameterizes the NP
amplitude with $\Delta I=0$. As explained above, the QCDF approach predicts
small values $\Delta \epsilon \approx 0$. In the following NP fits to
$B \to K\pi$ decays, we will therefore fix $\Delta \epsilon=0$ 
for simplicity, and vary the common values in the ranges
\begin{eqnarray}
&&  \epsilon_T = \epsilon_{3/2} = 0.23 \pm 0.06_{\rm flat} \pm 0.05_{\rm gauss}
     \,, \qquad
  \phi_T = \phi_{3/2} =  -0.13 \pm 0.11_{\rm flat} \,,
\label{eps:bbns}
\end{eqnarray}
 which have been determined by combining the 
 QCDF errors \cite{Beneke:2001ev} on the
 individual parameters (flat errors are combined linearly, 
 and the larger of the Gaussian errors is chosen).
As in the $B \to \phi K$ example, 
since the leading SM amplitudes are already penguin-suppressed, we expect
$r_0 \leq {\cal O}(1)$ and $\phi_0 \leq {\cal O}(\lambda)$. Generically, 
we now expect a sizeable direct CP asymmetry in $B^- \to \pi^- \bar K^0$
of the order $\lambda$. The experimental value for that asymmetry should
therefore be included in the fit and will essentially constrain 
the parameter combination $r_0 \, \sin\phi_0$.
On the other hand, using the power-counting
$\epsilon_i, q_C,\omega,\phi_i \sim \lambda$, 
a $\Delta I=0$ NP operator
does not contribute to the critical observables $A_I$ and $\Delta A_{CP}$
in (\ref{CritObs}) at order $\lambda$, either. 
As explained in \cite{Fleischer:2007mq} and
references therein, these observables are sensitive to $\Delta I=1$
operators which, in the SM, are represented by electroweak penguin operators.

As a result, the NP fit with $\Delta I=0$ contributions generally
leads to a bad description of the experimental data, except for
certain fine-tuned parameter combinations\footnote{
Notice, that contrary to the $B \to \phi K$ and $B \to J/\psi K$
analyses, we cannot exploit reparameterization invariance here, because
we decided to constrain certain hadronic input values from QCDF.
As a consequence, the fit results will explicitly depend on the
value of the NP weak phase.} with small
NP phase $\theta_W$ and 
unreasonably large values for the amplitude normalization factor $P$.
To avoid such fine-tuned scenarios, we consider some particular
examples with fixed NP phase $\theta_W$, see Table~\ref{tab:tabpiK0}.
We thus confirm on a quantitative level that $\Delta I=0$ NP contributions
alone cannot resolve the $B \to K\pi$ ``puzzles''.


  \begin{table}[t!bpt!]
  \begin{center}
  \parbox{0.9\textwidth}{
  \caption{\label{tab:tabpiK0} \small
  Fit to $\Delta I=0$ NP contribution in $B \to K\pi$.
  We show the $1\sigma$~confidence levels, assuming
  $\Delta \epsilon=0$
  (with $\epsilon_T \, e^{i\phi_T}=\epsilon_{3/2} \, e^{i\phi_{3/2}}$ varied
  according to their QCDF ranges, see text), $\epsilon_a=0$ and with $q \,
  e^{i\omega}$ and $q_C \, e^{i \omega_C}$
  varied according to (\ref{q},\ref{qc}) from \cite{Beneke:2001ev}.
  }
  \vspace{0.8em}
  }
\begin{tabular}{|c||c|c|c|}
\hline
\multicolumn{1}{|c|}{$\theta_W$} &  $|r_0|$
  &  $ \tan \phi_0$ & $\chi^2/{\rm d.o.f.}$ \\
\hline \hline
   $5\pi/6$
& [0.31 to 0.43] &  [0.00 to 0.03] & 14.9/6
\\
   $2\pi/3$
& [0.23 to 0.35] &  [0.01 to 0.06] & 17.9/6
\\
   $\pi-\gamma_{\rm SM}$
  &  [0.22 to 0.34]  &  [0.01 to 0.07]
& 18.5/6 \\
   $\pi/3$
& [0.23 to 0.50] &  [0.06 to 0.15] & 24.6/6
\\
   $\pi/6$
& [0.15 to 0.68] &  [0.21 to 0.54] & 34.4/6
\\
\hline
\end{tabular}
  \end{center}
  \end{table}


\subsubsection{New physics with $\Delta I=0,1$ 
 ($b \to s u\bar u$ or $b \to s d\bar d$)}

\begin{figure}[p!bpt!]
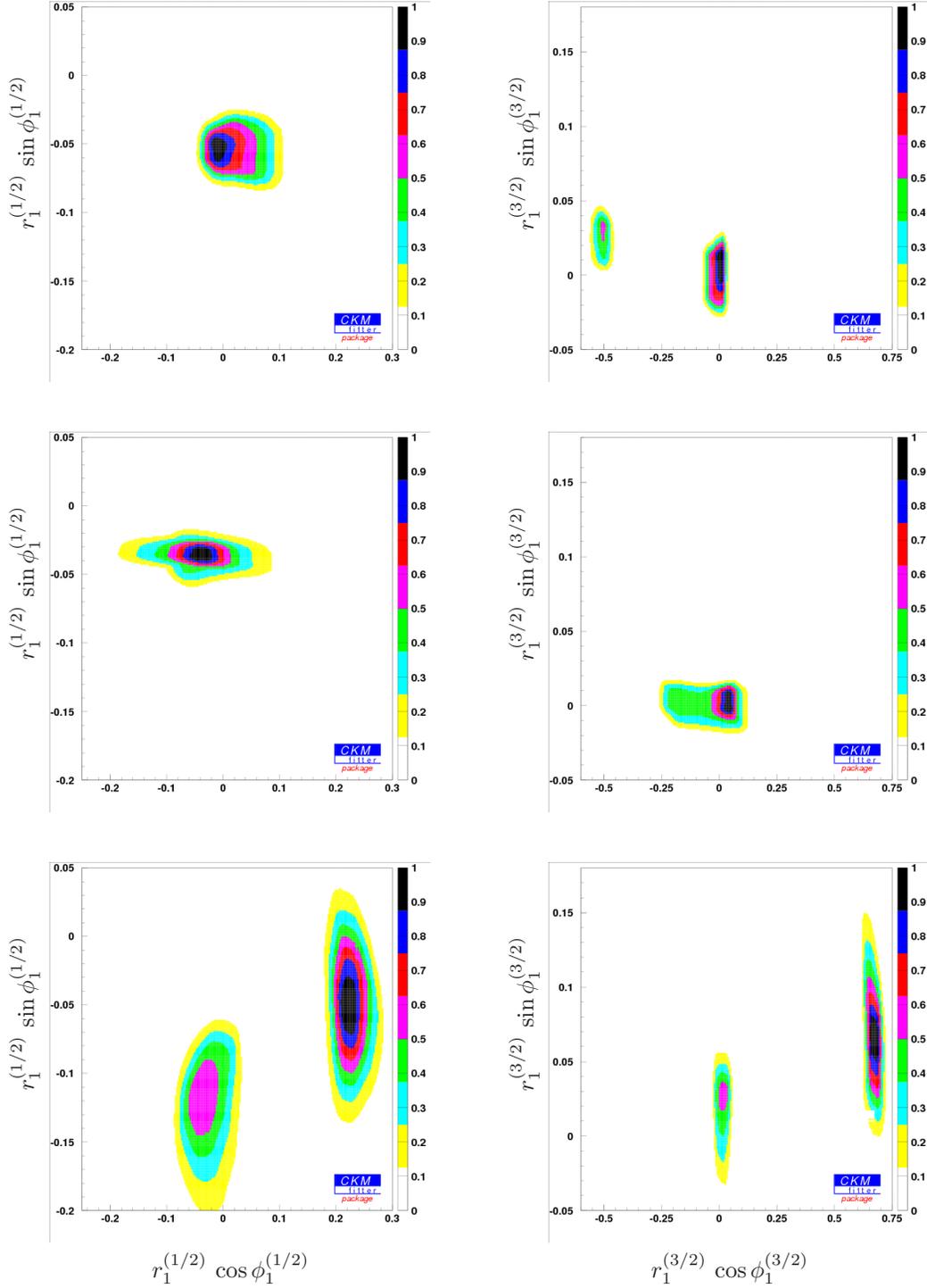

\begin{center}
\parbox{0.4\textwidth}{
\rotatebox{90}{\hspace{0.14\textwidth}\footnotesize $r_1^{(1/2)} \, \sin\phi_1^{(1/2)}$}
\includegraphics[width=0.35\textwidth]{r12c_fix30.eps2}\\
}
\qquad
\parbox{0.4\textwidth}{
\rotatebox{90}{\hspace{0.14\textwidth}\footnotesize $r_1^{(3/2)} \, \sin\phi_1^{(3/2)}$}
\includegraphics[width=0.35\textwidth]{r32c_fix30.eps2}\\
}
\\[0.5em]
\parbox{0.4\textwidth}{
\rotatebox{90}{\hspace{0.14\textwidth}\footnotesize $r_1^{(1/2)} \, \sin\phi_1^{(1/2)}$}
\includegraphics[width=0.35\textwidth]{r12c_fixgamma.eps2}\\
}
\qquad
\parbox{0.4\textwidth}{
\rotatebox{90}{\hspace{0.14\textwidth}\footnotesize $r_1^{(3/2)} \, \sin\phi_1^{(3/2)}$}
\includegraphics[width=0.35\textwidth]{r32c_fixgamma.eps2}\\
}
\\[0.5em]
\parbox{0.4\textwidth}{
\rotatebox{90}{\hspace{0.14\textwidth}\footnotesize $r_1^{(1/2)} \, \sin\phi_1^{(1/2)}$}
\includegraphics[width=0.35\textwidth]{r12c_fix150.eps2}\\
\centerline{\footnotesize  \ $r_1^{(1/2)} \, \cos\phi_1^{(1/2)}$}}
\qquad
\parbox{0.4\textwidth}{
\rotatebox{90}{\hspace{0.14\textwidth}\footnotesize $r_1^{(3/2)} \, \sin\phi_1^{(3/2)}$}
\includegraphics[width=0.35\textwidth]{r32c_fix150.eps2}\\
\centerline{\footnotesize  \ $r_1^{(3/2)} \, \cos\phi_1^{(3/2)}$}}
\vspace{0.8em}
\parbox{0.9\textwidth}{\caption{\small \label{fig:r01piK}
Fit results for $\Delta I=1$ NP contributions
$r_1^{1/2} \, e^{i\phi_1^{1/2}}$ (left) and
$r_1^{3/2} \, e^{i\phi_1^{3/2}}$ (right),
with  $\epsilon_a=0$, $\Delta \epsilon=0$ and
 ${\cal A}_{\rm CP}^{\rm dir}(\pi^- \bar K^0)=0$, see also text.
The plots in the upper row refer to a weak phase $\theta_W = 5\pi/6$,
the ones in the middle row to $\theta_W=\pi-\gamma$, and the lower
ones to $\theta_W=\pi/6$.
}}
\end{center}
\end{figure}

New physics contributions with $\Delta I=1$ induce two new isospin amplitudes
$$
  r_1^{(1/2)} \, e^{i \theta_W} \, e^{i \phi_{1}^{(1/2)}} \, P \,, \quad
\mbox{and} \quad
r_1^{(3/2)} \, e^{i \theta_W} \, e^{i \phi_{1}^{(3/2)}} \, P \,,
$$
corresponding to final $|K\pi\rangle$ state with $I=1/2$ or $I=3/2$.
Using the connection between (\ref{piKSM}) and isospin amplitudes (see e.g.\
\cite{Feldmann:2004mg}), we obtain
(again within our approximation)
\begin{eqnarray}
  {\cal A}(B^- \to \pi^- \bar K^0) &\simeq &
    {P} \Big( 1 + \left[r_0 \, e^{i\phi_{0}} 
 +  r_1^{(1/2)} \, \, e^{i \phi_{1}^{(1/2)}} 
 + r_1^{(3/2)} \,  e^{i \phi_{1}^{(3/2)}} \right] e^{i \theta_W}
  \Big) \,,
\nonumber \\
 -\sqrt2\, {\cal A}(B^- \to \pi^0  K^-) &\simeq &
    {P} \, \Big( 1 + r_0 \, e^{i\phi_{0}} \, e^{i\theta_W} 
     - \epsilon_{3/2} \, e^{i \phi_{3/2}} \left( e^{-i\gamma} - q \, e^{i\omega} \right)
\cr
&&  \qquad {} + \left[ r_1^{(1/2)} \, \, e^{i \phi_{1}^{(1/2)}} 
 -2 r_1^{(3/2)} \,  e^{i \phi_{1}^{(3/2)}} \right]   e^{i \theta_W}
\Big)
\,, \nonumber \\
 - {\cal A}(\bar B_d \to \pi^+  K^-) &\simeq&
   {P} \, \Big( 1 + r_0 \, e^{i\phi_{0}} \, e^{i \theta_W}
   - \epsilon_{T} \, e^{i \phi_T} \left( e^{-i\gamma} - q_C \, e^{i \omega_C} \right) \,
\cr
&& \qquad {} - \left[ r_1^{(1/2)} \, \, e^{i \phi_{1}^{(1/2)}} 
 + r_1^{(3/2)} \,  e^{i \phi_{1}^{(3/2)}} \right]  e^{i \theta_W}
\Big) \,.
\end{eqnarray}
In order to reduce the number of free parameters in the fit, and to
avoid unphysical solutions, we apply
additional assumptions/approximations:
\begin{itemize}

 \item Following the experimental observation,
       we force the direct CP asymmetry in $B^- \to \pi^-\bar K^0$ 
       to vanish identically, which yields the relation
$$
   r_0 \, e^{i\phi_{0}} 
 + r_1^{(1/2)} \, \, e^{i \phi_{1}^{(1/2)}} 
 + r_1^{(3/2)} \,  e^{i \phi_{1}^{(3/2)}}   = 0 \,,
$$
which we use to eliminate the parameters $r_0$ and $\phi_0$.
This effectively implies that we deal with  
a $b \to s u \bar u$ operator which does not contribute
to $B^- \to \pi^- \bar K^0$ in the naive factorization approximation.

\item 
Again, we assume the SM contributions to the
amplitude parameters $\epsilon_T $ and $\epsilon_{3/2}$
to lie within the QCDF ranges, see (\ref{eps:bbns}).
\end{itemize}
In Fig.~\ref{fig:r01piK} we display the results for the NP parameters
$r_1^{(1/2)} \, e^{i \phi_1^{(1/2)}}$ and $r_1^{(3/2)} \, e^{i \phi_1^{(3/2)}}$
in the complex plane, for different values of the NP weak phase $\theta_W$. 
The corresponding $1\sigma$ ranges are collected in Table~\ref{tab:tabpiK1}.
 \begin{table}[t!bpt!]
  \begin{center}
  \parbox{0.9\textwidth}{
  \caption{\label{tab:tabpiK1} \small
  Same as Table~\ref{tab:tabpiK0} for the fit with $\Delta I=0,1$ NP contribution in $B \to K\pi$.
  }
  \vspace{0.8em}
  }
\begin{tabular}{|c||c|c|c|c|c|}
\hline
\multicolumn{1}{|c|}{$\theta_W$} 
  &  $|r_1^{(1/2)}|$ &  $ \tan\phi_1^{(1/2)} $
  &  $|r_1^{(3/2)}|$ &  $ \tan\phi_1^{(3/2)} $ & $\chi^2/{\rm d.o.f.}$ \\
\hline \hline
   $5\pi/6$
 & [0.04 to 0.08] &  [\phantom{-}0.06 to \phantom{-}0.08] 
 & [0.00 to 0.04] &  unconstr. 
 & 4.3/3
\\
   $2\pi/3$
 & [0.03 to 0.07] &  [-2.65 to -0.51] 
 & [0.00 to 0.05] &  unconstr.
 & 3.5/3
\\
   $\pi-\gamma_{\rm SM}$
 &  [0.03 to 0.09]  &  [-9.89 to \phantom{-}0.38]
 &  [0.00 to 0.07]  & unconstr.
 & 2.9/3
 \\
   $\pi/3$
 & [0.04 to 0.11] &  [-16.4 to -0.42]
 & [0.41 to 0.51] &  unconstr.
 & 0.4/3
\\
   $\pi/6$
 & [0.20 to 0.26] &  [-0.41 to -0.03]
 & [0.65 to 0.70] &  unconstr. 
 & 1.7/3
\\
\hline
\end{tabular}
  \end{center}
  \end{table}
The resulting central values for the observables in the case $\theta_W=\pi-\gamma_{\rm SM}$
are listed in the last column of Table~\ref{tab:bestfit}.
We observe that the fit depends on the value of the NP weak phase $\theta_W$
in an essential way. In particular, depending on whether $\theta_W$ is less or
greater than $\pi/2$, we encounter disjunct regions in parameter space.
One of the regions always corresponds to relatively small values 
of $r_1^{(1/2,3/2)} \lesssim 10\%$, whereas for values of $\theta_W$ close
to $0$ or $\pi$ solutions with $r_1^{(1/2,3/2)}$ as large 50\% are possible.

\clearpage

\section{Conclusions} 

\label{sec:concl}

To date, flavour physics is evolving from the 
$B$\/-factory era to the LHC era. While the former has led to an enormously
successful confirmation of the CKM mechanism in the SM, the latter is expected
to reveal direct and indirect signs for physics beyond the SM with interesting
interplay between high-$p_T$ and flavour physics \cite{LHCreport,Isidori:2008qp,Browder:2008em}.
In this context, a crucial task is to constrain the flavour structure of
NP models, manifesting itself in rare quark and lepton decays and 
production and decay of new flavoured particles. 

While within concrete NP models the chiral, flavour and colour structure 
of new operators could be completely specified, the present work pursues
a model-independent approach. Assuming the dominance of an individual
NP operator, the analysis of $B \to J/\psi K$,
$B \to \phi K$ and $B \to K \pi$ observables allows us 
to infer semi-quantitative information about the relative size 
of NP contributions to 
$b \to s \, c\bar c$, $b \to s \, s\bar s$, $b \to s \, d\bar d$,
and $b \to s \, u\bar u$ operators. 
The main conclusions to be drawn are:
\begin{itemize}
 \item From the comparison of isospin-averaged $B \to J/\psi K$ and 
       $B \to \phi K$ observables we find that -- after correcting
       for relative penguin, phase-space and normalization factors --
       NP contributions to $b \to s c\bar c$ and $b \to s s\bar s$ operators
       may be of similar size (order 10\% relative to a SM tree operator).

 \item In a scenario, where $b \to s d \bar d$ is the \emph{only} source for
  NP contributions in $B \to \pi K$ observables, while the SM contributions
  are estimated in QCD factorization, one cannot
  simultaneously explain the individual CP asymmetries. 
  In particular, the experimental value for $A_{\rm CP}(\pi^+ K^-)$, which 
  does not receive leading
  NP contributions from $b \to s d \bar d$, cannot be
  reproduced in a scenario with negative strong phase $\phi_T$.

  Moreover, the small direct CP asymmetry for $B^- \to \pi^-\bar K^0$
  requires the matrix element of a $b \to s d \bar d$ NP operator to 
  have either a small coefficient or a small phase.

 \item This leaves the $b\to s u\bar u$ operators, which correlate 
       isospin-violating observables in $B \to J/\psi K$ and
      $B \to K\pi$ decays, and may be even somewhat larger (order 20\%
      relative to a SM tree operator) than the $b\to s c\bar c$ and
      $b \to s s\bar s$ operators.
 \end{itemize}
In all cases, in order to explain deviations from SM expectations
for CP asymmetries without fine-tuning of hadronic parameters
(see the discussion after (\ref{reparapprox})), 
we have to require non-trivial weak phases ($\theta_W\neq 0,\pi$), 
which could be due to NP, albeit the case $\theta_W = \pi -\gamma_{\rm SM}$ 
is always allowed, too.
Consequently, our findings are still compatible with a SM scenario
where non-factorizable QCD dynamics in matrix elements of
sub-leading operators is unexpectedly large.

In the future, an improvement of experimental accuracy, in particular on the isospin-violating observables, could lead to even more 
interesting constraints on the relative importance
of different $b \to s q\bar q$ operators and their interpretation within
particular NP models with 
MFV \cite{D'Ambrosio:2002ex,Ciuchini:1998xy,Buras:2000dm} or beyond 
(see e.g.\ \cite{Agashe:2005hk,Feldmann:2006jk,Lunghi:2007ak,Fitzpatrick:2007sa,Davidson:2007si}).

\section*{Acknowledgements}

This work is supported by the German Ministry of Research
(BMBF) under contract No.~05HT6PSA. T.F.\ acknowledges financial
support by the Cluster of Excellence ``Origin and Structure
of the Universe'' during his stay at the Technical University
Munich.


\end{document}